\documentstyle[12pt,epsf]{article}
\newlength{\dinwidth}               \newlength{\dinmargin}
\begin{document}
\newcommand{\jpsi}     {\mbox{J/$\psi$}}
\newcommand{\jpsitoee} {\mbox{J/$\psi \rightarrow e^{+}e^{-} $}}
\newcommand{\jpsitomm} {\mbox{J/$\psi \rightarrow \mu^{+}\mu^{-} $}}
\newcommand{\ppsi}     {\mbox{$\psi (3685)$}}
\newcommand{\gev}      {\mbox{${\rm GeV}$}}
\newcommand{\pip}      {\mbox{${\rm \pi^{+}}$}}
\newcommand{\pim}      {\mbox{${\rm \pi^{-}}$}}
\newcommand{\ppm}      {$\pm$}
\newcommand{\ee}       {\mbox{${\rm e^{+}e^{-}}$}}
\newcommand{\mm}       {\mbox{${\rm \mu^{+}\mu^{-}}$}}
\newcommand{\mee}      {\mbox{${\rm M_{ee}}$}}
\newcommand{\mmm}      {\mbox{${\rm M_{\mu\mu}}$}}
\newcommand{\wgp}      {\mbox{${\rm W_{\gamma p}}$~}}
\newcommand{\qsquare}  {\mbox{$Q^2$}}
\newcommand{\ptsq}     {\mbox{${p_{T}^{2}}$}}
\newcommand{\micron}   {\mbox{${\rm \mu m}$}}
\newcommand{\ccbar}    {\mbox{${\rm c\bar c}$}}
%  choose/remove the section you want placing '%' before \include:
\title{
{\bf     Measurement of the Cross Section for the
Reaction {\boldmath {$\rm \gamma p\rightarrow \jpsi~p$}} with the
ZEUS Detector at HERA}
\author
{\rm  ZEUS Collaboration \\}
}
\date{ }
\maketitle
\vspace{5cm}

\begin{abstract}
This paper reports the cross section measurements for the process ep
$\rightarrow$ e~\jpsi~p for $Q^2 < 4$ GeV$^2$ at $\sqrt s = 296$ GeV,
based on an integrated luminosity of about 0.5~pb$^{-1}$, using the ZEUS
detector. The \jpsi~was detected in its $\rm e^{+}e^{-}$ and $\rm
\mu^{+}\mu^{-}$ decay modes.  The photoproduction cross section was
measured to be $52^{ \ +7}_{-12}\pm 10$~nb at an average $\gamma$p centre
of mass energy of 67 GeV and $71^{+13}_{-20}\pm 12$~nb at 114 GeV.  The
significant rise of the cross section compared to lower energy
measurements is not in agreement with VDM models, but can be described by
QCD inspired models if a rise in the gluon momentum density at low $x$ in
the proton is assumed.
\end{abstract}

\vspace{-20cm}
\begin{flushleft}
\tt DESY 95-052 \\
March 1995 \\
\end{flushleft}

\setcounter{page}{0}
\thispagestyle{empty}

\newpage

%   02/03/95 503101048  MEMBER NAME  AUTH022  (ZEUS)     M  TEX
%
\def\3{\ss}
\parindent 0cm
\footnotesize
\renewcommand{\thepage}{\Roman{page}}
\begin{center}
\begin{large}
The ZEUS Collaboration
\end{large}
\end{center}
M.~Derrick, D.~Krakauer, S.~Magill, D.~Mikunas, B.~Musgrave,
J.~Repond, R.~Stanek, R.L.~Talaga, H.~Zhang \\
{\it Argonne National Laboratory, Argonne, IL, USA}~$^{p}$\\[6pt]
R.~Ayad$^1$, G.~Bari, M.~Basile,
L.~Bellagamba, D.~Boscherini, A.~Bruni, G.~Bruni, P.~Bruni, G.~Cara
Romeo, G.~Castellini$^{2}$, M.~Chiarini,
L.~Cifarelli$^{3}$, F.~Cindolo, A.~Contin, M.~Corradi,
I.~Gialas$^{4}$,
P.~Giusti, G.~Iacobucci, G.~Laurenti, G.~Levi, A.~Margotti,
T.~Massam, R.~Nania, C.~Nemoz, \\
F.~Palmonari, A.~Polini, G.~Sartorelli, R.~Timellini, Y.~Zamora
Garcia$^{1}$,
A.~Zichichi \\
{\it University and INFN Bologna, Bologna, Italy}~$^{f}$
% ...
\\[6pt]
\clearpage
\hspace*{1mm}
$^{ 1}$ supported by Worldlab, Lausanne, Switzerland \\
\hspace*{1mm}
$^{ 2}$ also at IROE Florence, Italy  \\
\hspace*{1mm}
$^{ 3}$ now at Univ. of Salerno and INFN Napoli, Italy  \\
\hspace*{1mm}
$^{ 4}$ supported by EU HCM contract ERB-CHRX-CT93-0376 \\
\hspace*{1mm}
% ...
$^{16}$ now with OPAL Collaboration, Faculty of Physics at Univ. of
        Freiburg \\
% ...

\begin{tabular}{lp{15cm}}
$^{a}$ &supported by the Natural Sciences and Engineering Research
         Council of Canada (NSERC) \\
$^{b}$ &supported by the FCAR of Qu\'ebec, Canada\\
$^{c}$ &supported by the German Federal Ministry for Research and
         Technology (BMFT)\\
$^{d}$ &supported by the MINERVA Gesellschaft f\"ur Forschung GmbH,
         and by the Israel Academy of Science \\
$^{e}$ &supported by the German Israeli Foundation, and
         by the Israel Academy of Science \\
$^{f}$ &supported by the Italian National Institute for Nuclear Physics
         (INFN) \\
$^{g}$ &supported by the Japanese Ministry of Education, Science and
         Culture (the Monbusho)
         and its grants for Scientific Research\\
$^{h}$ &supported by the Korean Ministry of Education and Korea Science
         and Engineering Foundation \\
$^{i}$ &supported by the Netherlands Foundation for Research on Matter
         (FOM)\\
$^{j}$ &supported by the Polish State Committee for Scientific Research
         (grant No. SPB/P3/202/93) and the Foundation for Polish-
         German Collaboration (proj. No. 506/92) \\
$^{k}$ &supported by the Polish State Committee for Scientific
         Research (grant No. PB 861/2/91 and No. 2 2372 9102,
         grant No. PB 2 2376 9102 and No. PB 2 0092 9101) \\
$^{l}$ &partially supported by the German Federal Ministry for
         Research and Technology (BMFT) \\
$^{m}$ &supported by the German Federal Ministry for Research and
         Technology (BMFT), the Volkswagen Foundation, and the Deutsche
         Forschungsgemeinschaft \\
$^{n}$ &supported by the Spanish Ministry of Education and Science
         through funds provided by CICYT \\
$^{o}$ &supported by the Particle Physics and Astronomy Research
        Council \\
$^{p}$ &supported by the US Department of Energy \\
$^{q}$ &supported by the US National Science Foundation
\end{tabular}

\newpage
\pagenumbering{arabic}
\setcounter{page}{1}
\normalsize

\section{\bf Introduction}
\label{s:intro}

\begin{figure}[hptb]
\unitlength1cm  \begin{picture}(8,15.0)
\includegraphics{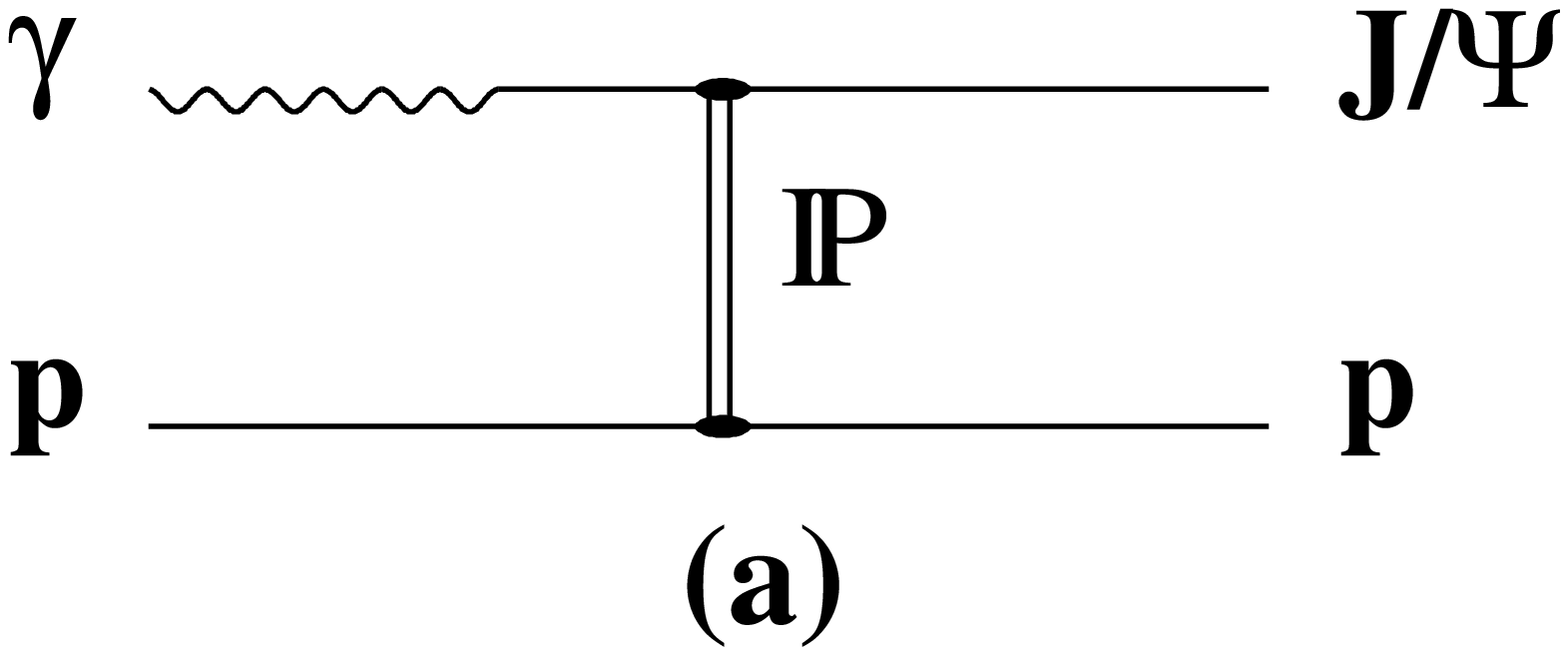}
\includegraphics{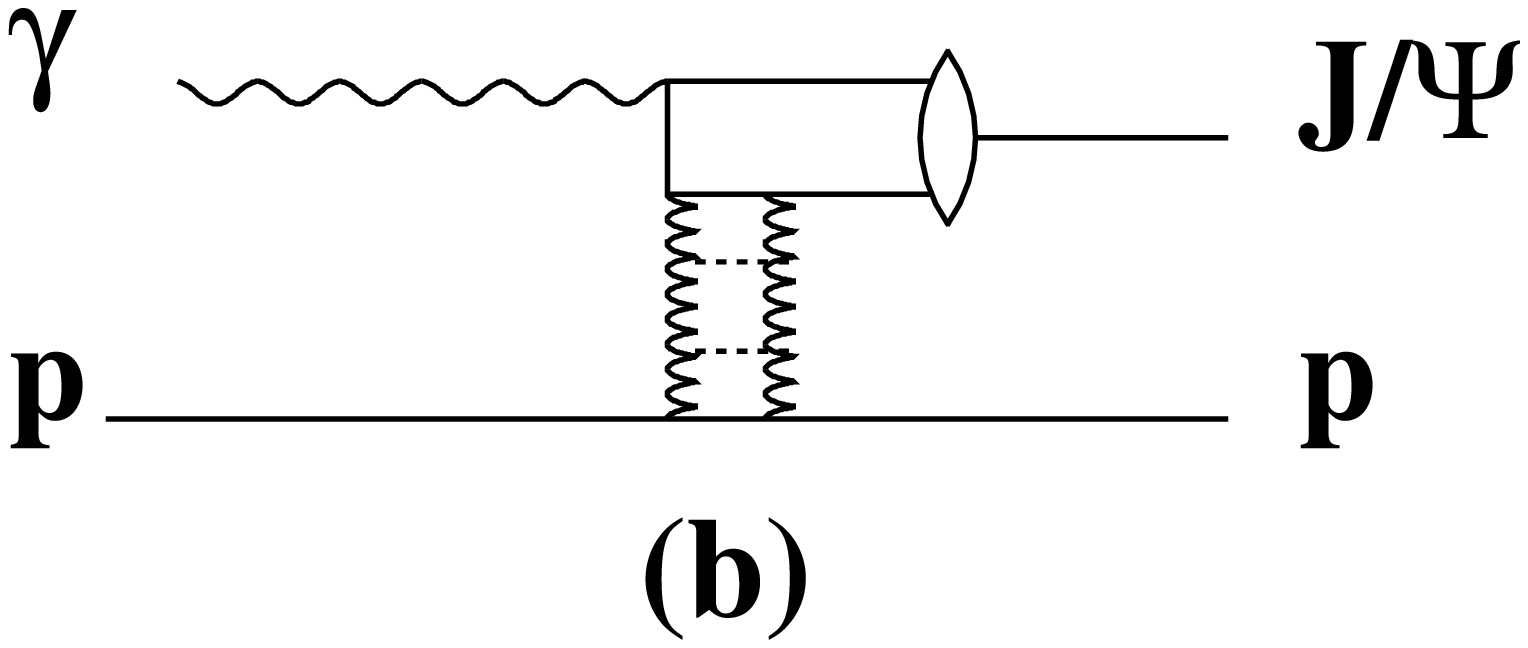}
%\special{psfile=/canada/zurich/lim/jpsi/elastic/fig1_c.ps angle=0
%vscale=50 hscale=30 voffset=-200 hoffset=150}
\end{picture}
\caption{A schematic diagram of elastic \jpsi~production
according to (a)VDM with a pomeron exchange and (b) QCD-inspired models with
the exchange of a gluon ladder.}
\label{FEYNM}
\vspace{-0.2cm}
\end{figure}

Elastic photoproduction of \jpsi~($\gamma$p$\rightarrow \jpsi$p)
is particularly interesting as the
production cross section can be calculated as a function of
the $\gamma$p centre of mass (c.m.) energy, $W$, both using the Vector
Dominance Model\cite{VDM}~(VDM)
extended to the heavy quarks \cite{DANDL,GLM12,TERRON}, and,
because of the large value of the \jpsi~mass ($M_{\rm{J}/\psi}$), with
QCD inspired models\cite{RYSKIN, NIKZAK}.
Figures~\ref{FEYNM}a and \ref{FEYNM}b show the elastic \jpsi~photoproduction
mechanisms according
to VDM and QCD inspired models, respectively.
A characteristic of these QCD inspired
models is that the cross section is proportional to the square
of the proton's gluon momentum density.
The typical $x$ range probed here, where
$x~(\simeq M_{\rm{J}/\psi}^2/W^2)$
is the fraction of the proton momentum carried by the gluon,
is approximately $5\times 10^{-3}$ to $5\times 10^{-4}$,
corresponding to a $W$ range between 40 and 140 GeV.
%The QCD scale at which
%the gluon distribution is probed is $\sim M^{2}_{\rm{J}/\psi} / 4 $
%\cite{RYSKIN}.
For gluon distributions increasing at low $x$, the QCD approaches
predict much higher cross sections
than those from the VDM in this $W$ region.

Previous measurements of the photoproduction of \jpsi~are
available in the $W$ range between 4 GeV and 28 GeV\cite{AL_xsect1,AL_xsect2}.
However, some results include elastic as well as other
production mechanisms of the \jpsi. The H1 experiment
has also reported\cite{H1M}~a \jpsi~photoproduction cross section
measurement at an average c.m. energy of 90 GeV, that
contains an unknown contribution from inelastic \jpsi~photoproduction.

This paper reports the measurements of photoproduction
cross sections of elastically produced \jpsi's in the reaction
$\rm ep \rightarrow e~\jpsi~p$, followed by \jpsitoee or \jpsitomm,
with the ZEUS detector. The present data sample contains
events with $\qsquare <$ 4 GeV$^2$, with a median $\qsquare\sim 10^{-3}$
GeV$^2$, in the $W$ range between 40 and 140 GeV.
Neither the scattered electron nor the proton is detected in this
measurement.
The contribution of \jpsi~production where the proton diffractively
dissociated is subtracted to obtain the elastic photoproduction cross section.

\section{\bf HERA }

The data presented were collected during the 1993 running period of HERA
and represent an integrated luminosity of about $\rm 0.5~pb^{-1}$.
HERA operated in 1993 with an electron beam energy of 26.7~GeV and a proton
beam energy of 820~GeV. A total of 84 colliding
electron and proton bunches was used; in addition
10 electron and 6 proton unpaired bunches
were used for background studies. The time between
bunch crossings at HERA is 96 ns. A typical instantaneous luminosity of
$\rm \sim 6\times 10^{29}cm^{-2}s^{-1}$ was delivered.

\section{\bf The ZEUS detector}

The ZEUS detector\cite{ZEUS} is a hermetic, general purpose magnetic detector
with a tracking region surrounded by a high resolution calorimeter followed
in turn by a backing calorimeter and the
muon detector.
A short description of the components relevant to this analysis is
given here. They are the vertex detector\cite{VXD},
the central-tracking detector\cite{CTD}, the uranium-scintillator
calorimeter\cite{CAL} and the barrel and rear muon
detectors\cite{BRMUON}.

Charged particles are measured by the ZEUS inner tracking detectors,
which operate in a
magnetic field of 1.43 T provided by a thin superconducting coil.
Immediately surrounding the beampipe is the vertex detector (VXD)
consisting of 120 radial cells, each with 12 sense wires. It
uses a slow drift velocity gas and the presently achieved resolution
in the $XY$\footnote{
The ZEUS detector uses a right-handed coordinate system where the Z axis
points in the direction of the proton beam (forward direction)
and the X axis is horizontal, pointing towards the centre of HERA,
with the nominal Interaction Point (IP) at (0,0,0).}
plane is 50 $\mu$m in the central region of a cell and 150
$\mu$m near the edges.
Surrounding the VXD is
the central tracking detector (CTD) which consists of 72 cylindrical
drift chamber layers, arranged in 9 `superlayers'.
With our present calibration of the chamber,
the resolution of the CTD is around $260~\mu$m.
The resolution in transverse momentum for tracks going through all
superlayers is
$\sigma (p_T) /p_T \approx  \sqrt{ (0.005)^2\, p_{T}^2 + (0.016)^2}$ where
$p_T$ is in GeV.
The single hit efficiency is greater than 95\%.
The efficiency for assigning hits to tracks depends on several factors:
very low
$p_T$ tracks suffer large systematic effects which reduce the
probability of hits being assigned to them, and the $45^{\circ}$ inclination of
the drift cells
also introduces an asymmetry between positive and
negative tracks. Nevertheless, the track reconstruction efficiency for
tracks with $p_T > 0.1$ GeV is greater than 95\%.
Using the combined data from the VXD and CTD,
resolutions of $0.4$ cm in $Z$ and $0.1$ cm in radius
in the $XY$ plane
are obtained for the primary vertex reconstruction.
{}From Gaussian fits to the $Z$ vertex
distribution, the r.m.s. spread is found to be $10.5$ cm, in agreement with
the expectation based on the HERA proton bunch length.

The high resolution
uranium-scintillator calorimeter (CAL)
%is used in the present ana\-ly\-sis
covers the
polar angle range between $2.2^{\circ} < \theta <
176.5^{\circ}$, where $\theta = 0^{\circ}$ is the proton beam direction. It
consists of three parts: the rear calorimeter (RCAL), covering the backward
pseudorapidity\footnote{Pseudorapidity is defined as $\rm \: \eta = -ln
[tan(\frac{\theta}{2})].$}
range ($-3.4 < \eta < -0.75$);
the barrel calorimeter (BCAL) covering the central region ($-0.75 < \eta <
1.1$); and the forward calorimeter (FCAL) covering the forward region ($1.1 <
\eta < 3.8$). The calorimeter parts are subdivided into towers which
in turn are subdivided longitudinally into electromagnetic (EMC)
and hadronic (HAC) sections. The sections are subdivided into cells,
each of which is viewed by two photomultiplier tubes.
Under test beam conditions the CAL has an energy resolution, in units of
GeV, of $\sigma_{E} = 0.35 \sqrt{E(\rm GeV)}$ for hadrons and
$\sigma_{E} = 0.18 \sqrt{E(\rm GeV)}$ for electrons.
The CAL also provides a time resolution of better than 1 ns for energy
deposits greater than 4.5 GeV, and this timing is used
for background rejection.

The muon detectors, placed outside the calorimeter, are also
divided into three sections, covering the forward, barrel and rear regions.
In the barrel and rear regions limited streamer tube (LST) chambers
before (inner) and after (outer) an 80~cm thick magnetized iron yoke are used.
Only the inner chambers of the barrel and rear muon detectors (BMUI and RMUI)
were used
for the present analysis. The BMUI and the RMUI cover the polar angles
between $34^{\circ}\le\theta\le 135^{\circ}$ and
$134^{\circ}\le\theta < 171^{\circ}$, respectively.
%The bottom octant has only the outer chamber.
Each chamber has 2 double
layers of LST; spatial resolutions of 1 and 3~mm have been obtained along the
direction of the tube axis and perpendicular to it, respectively.

A set of four scintillation counters (C5) in two planes interleaved with a
3~mm Pb foil immediately behind the RCAL at approximately Z = $-3$~m partially
surrounds the beampipe. The C5 counter measures the timing of both
beams and also tags events from proton-gas interactions. A vetowall (VETO),
consisting of two layers of scintillator on either side of an 87
cm thick iron wall centred at Z = $-$7.3 m, was also used
to tag and reject off-axis beam particles.

The ep luminosity was measured from the rate of
the Bethe-Heitler process ($ep \rightarrow e\gamma p$) by counting the final
state photons in the luminosity monitor.

\section{\bf Kinematics}

The kinematics of the elastic process ep $\rightarrow$ e~\jpsi~p are described
below.
The incoming and outgoing electron four-momenta are denoted by $k$ and $k'$,
respectively, while the four-momentum of the virtual photon is $q = k - k'$;
$\qsquare \equiv - q^2$.
If $P(P')$ is the incoming(outgoing) proton four-momentum, and
$P_{\rm{J}/\psi}$
is the \jpsi~four-momentum, then
the squared four-momentum transfer $t$ is defined as
$$t = (P-P')^2 = (q-P_{\rm{J}/\psi})^2. $$
For $\qsquare\simeq 0$,  $t \simeq -\ptsq_{\rm{J}/\psi}$,
where $p_{T\rm{J}\psi}$ is the transverse momentum carried by the \jpsi.
$W$ is given by $ W^2 = (P + q)^2$.
The Lorentz scalar $y$ is defined as ${P \cdot q}/{P \cdot k}$,
and can be approximated as:
$$y \simeq {(E-p_Z)}_{\rm{J}/\psi}/2E_e, \ \ \rm{and} \ \ \it W^{\rm 2 } = sy
\simeq \rm 4 \it E_eE_py \simeq \rm 2 \it (E-p_Z)_{\rm{J}/\psi} E_p,$$
where $E_e$ denotes the electron beam energy, $E_p$ is the proton beam energy,
$\sqrt{s}$ is the $ep$ c.m. energy, and
$E$ is the energy and $p_Z$ is the $Z$ component
of the momentum of the \jpsi.

\section{\bf Trigger and preselection}
\subsection{\bf Trigger}

The \jpsi~ was identified from its leptonic decay modes.
The momenta
of the decay leptons from the \jpsi's in the observed kinematic
range are low
($\sim$ 1.5 GeV). To trigger on these low momentum lepton
tracks in the high background environment of HERA, where the
typical background rate from beam-gas interactions exceeds 10 KHz,
requires a very selective trigger.

ZEUS uses a three level trigger scheme\cite{ZEUS}.
The first level trigger (FLT) is built as a deadtime-free pipeline.
The triggering on the leptons from the decay of the \jpsi~used
the CTD, CAL and the muon chambers and is described below.

Events with low momentum electrons were selected two ways by the FLT:
\begin{itemize}
\item
by requiring a minimum energy deposit of 660 MeV in the EMC. To reduce
background, a total CAL energy deposit greater than 2 GeV or
a total energy deposit in the FCAL, excluding the region adjacent to the
beampipe,
greater than 2.5 GeV were required. In addition, one to three track
segments in the innermost superlayer of the CTD were required,
\item
by requiring a minimum energy deposit of 464 MeV in REMC and any track
segment in the innermost CTD superlayer.
\end{itemize}
In both of these cases, events were vetoed if
the energy deposit in the FCAL region immediately surrounding
the beam pipe exceeded 3.75 GeV to reduce the background from
proton gas interactions.

The analysis of the muon decay mode used the inclusive muon triggers;
events with low momentum muons were selected by the FLT in one of two ways
by requiring:
\begin{itemize}
\item
hits in the RMUI accompanied by an energy deposit of at least 464 MeV in the
RCAL,
\item
hits in the BMUI accompanied by an
energy deposit of at least 464 MeV in the CAL.
\end{itemize}
The trigger used only the inner muon chambers to maximize the acceptance.

All four of these triggers also required at least one CTD track
segment pointing
towards the IP, along with the appropriate VETO and C5 signals
to ascertain that the event originated from the IP.

The second level trigger (SLT) reduced the beam related
background further by making use of the subnanosecond
time resolution of the CAL and by requiring that the energy
deposits in the CAL were in time with the bunch crossing.

The third level trigger (TLT) ran on a farm of
processors; an event was flagged by the TLT as a \jpsi~candidate
if either of the following criteria was satisfied:

$\bullet$ Electron Mode:
a fast electron identification was carried out by using information from
the CTD and CAL. Electrons were identified by an
energy deposit in the EMC of at least 90\% of the cluster energy, where a
cluster is defined as a number of contiguous cells with energy deposit.
The tracks from the CTD were extrapolated
towards the CAL and matched\footnote
{The distance of closest approach between
the extrapolated track and the centre of the cluster was required to be less
than 30 cm.}
with a cluster in order to
determine the energy deposited in the calorimeter by that track.
An event was accepted if it had a pair of oppositely charged tracks,
each associated with such an electron cluster and each with a momentum
exceeding 0.5 GeV and $p_T$ higher than 0.4 GeV. The track reconstruction
in the TLT used
the `Z-by-timing' information available from the first three
axial superlayers of the CTD.
Additional requirements imposed on the events to remove beam-gas interactions
were
$\Sigma_i p_{Zi}/\Sigma_i E_i \leq 0.94$
and $\Sigma_i(E_i-p_{Zi})\leq 100$ GeV, where
the sums are over all calorimeter cells.

$\bullet$ Muon Mode:
for an event with a muon FLT trigger, a match between
track segments
from the CTD and the inner muon chambers was required, with energy deposits
compatible with those from a minimum ionizing particle in the calorimeter
EMC and HAC sections. A track with a minimum $p_T$ of 1 GeV  in the
barrel region or a minimum momentum of 1 GeV in the rear region was
flagged as an inclusive muon trigger and satisfied the TLT requirement.

\subsection{Preselection}
All events satisfying the TLT criteria were reconstructed offline
where the more refined information from the CTD was used for tracking.
The event samples were then divided into
an electron sample based on kinematics and track-CAL cluster
matching, and
a muon sample based on muon identification by the muon chambers.

$\bullet$ Electron Mode:
events with two oppositely charged tracks, each with $ p_T \geq 0.5$ GeV,
where one of the two had a minimum momentum of 1 GeV, were selected.
The tracks,
which were required to originate from the event vertex, were matched to a CAL
cluster as in the TLT, and were selected if the invariant $\ee~$mass was
between 2 and 4 GeV. This data sample contained
2021 events from a total luminosity of 486~nb$^{-1}$.

$\bullet$ Muon Mode: Cosmic ray events, the biggest background in
the muon sample, were substantially reduced
by using the time difference
between the upper and the lower halves of the calorimeter.
Again, events with
two oppositely charged tracks, each with $p_T \geq 0.5$ GeV,
originating from a vertex, were retained. A further reduction in beam gas
contamination was achieved by requiring $\Sigma_i p_{Zi}/\Sigma E_i \leq 0.96$,
where the sum was over all calorimeter cells.
This sample contained 456 events from a total
luminosity of 490~nb$^{-1}$.

\section{\bf Analysis}
\subsection{\bf The {\boldmath \jpsi} signal}
In order to ensure a high quality of track reconstruction,
for both data samples only tracks within the pseudorapidity
range $ | \eta | < 1.5 $ were considered. In addition, the total
energy in the CAL, apart from the
energy deposited by the two leptons from the \jpsi~candidate, was
required to be less than 1.0 GeV.
This criterion was imposed to reject
inelastic events. Only events where the acceptance was high
(the $W$ range between 40 and 140 GeV) were retained.

\begin{figure}[hptb]
\unitlength1cm  \begin{picture}(8,18.0)
\includegraphics{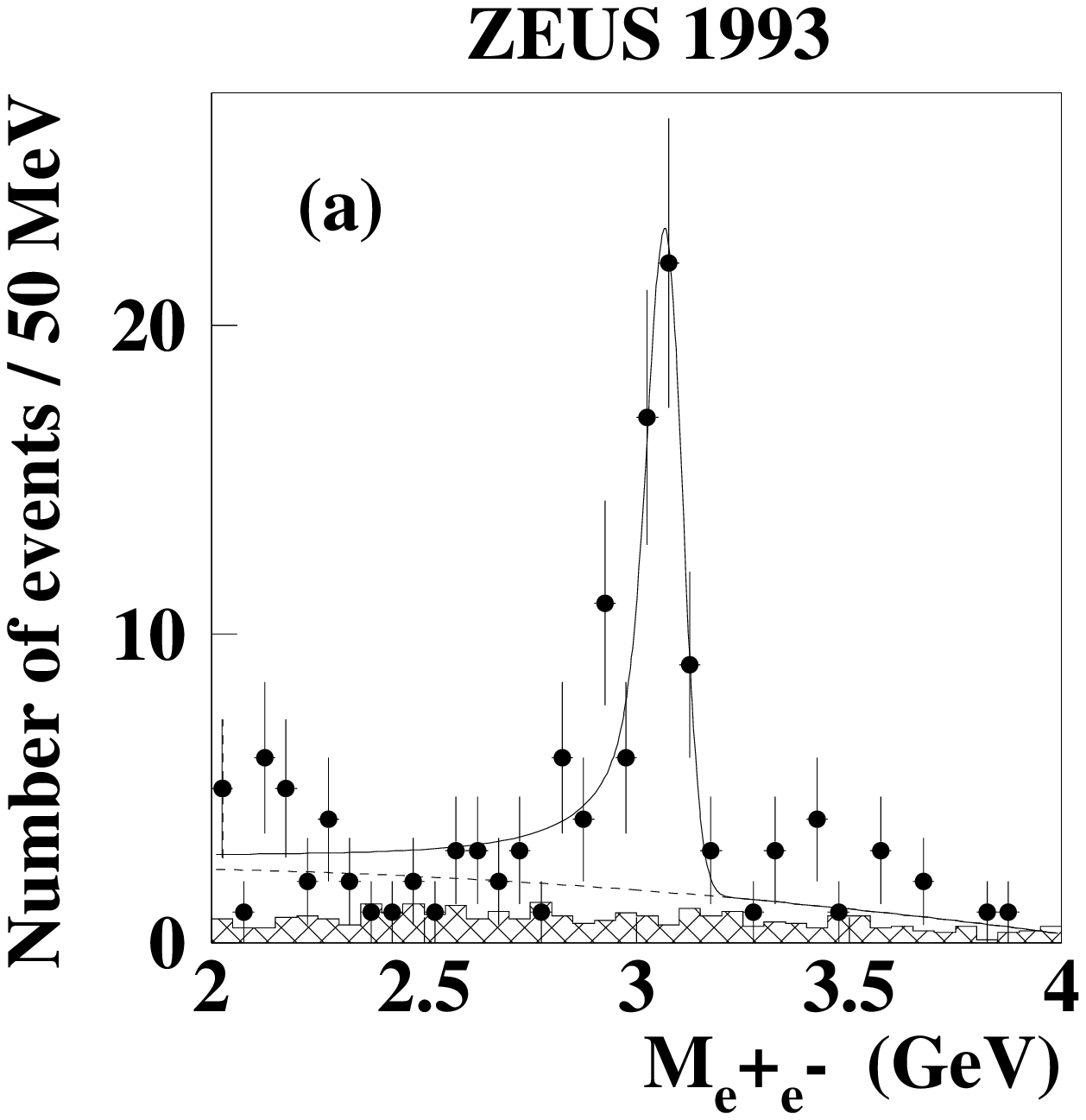}
\includegraphics{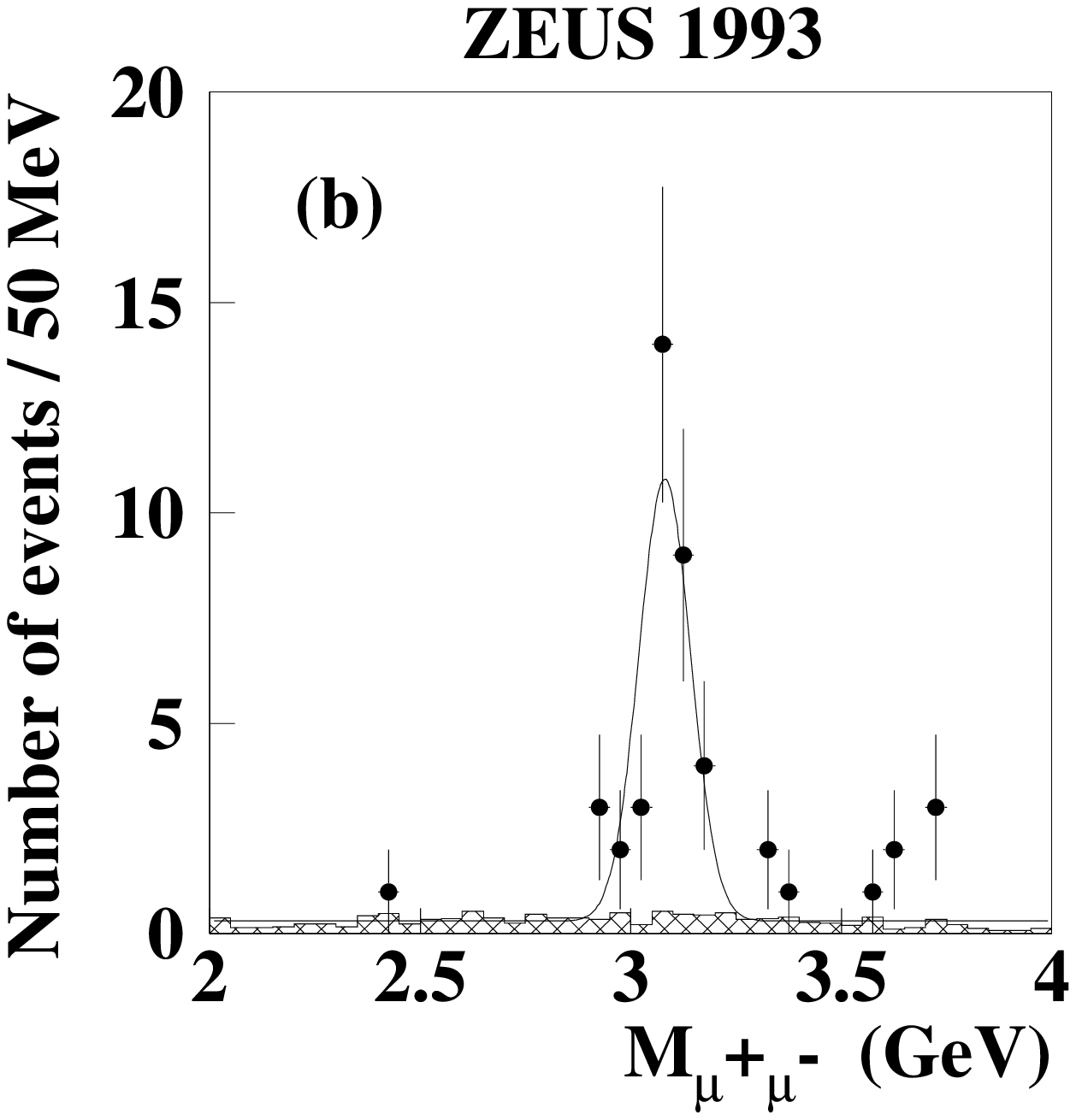}
\end{picture}
\caption{(a) The reconstructed \ee~invariant mass spectrum for
40 $\le W \le$ 140 GeV. The solid circles
represent the data; the solid
line indicates a fit to the data
with the convolution of a Gaussian and a bremsstrahlung
function; the dotted line represents a quadratic polynomial
parametrising the background; the shaded area is the expected
contribution from the two-photon background. (b) The
reconstructed \mm~invariant mass spectrum for 40 $\le W \le$ 140 GeV.
The solid circles represent the
data; the solid line indicates a fit with
a Gaussian and a flat background. The shaded area shows the flat
background expected from the two-photon process.}
\label{effwgp}
\vspace{-0.2cm}
\end{figure}

$\bullet$ Electron decay mode:
a total of 136 events satisfied these criteria with 72 events in the
invariant
mass range {\mbox {2.85 - 3.25}} GeV.
The resulting $\ee~$invariant mass distribution ($M_{\ee}$)
is shown in Fig.~\ref{effwgp}a. A clear peak
is visible at the \jpsi~mass. The asymmetric shape
is attributed to energy loss by the bremsstrahlung process
in the material encountered by the decay electrons.
The same asymmetry was observed in the reconstructed events from the Monte
Carlo study described below.
A maximum likelihood fit to the $\ee~$mass spectrum was
performed using a Gaussian shape convoluted with a
bremsstrahlung function to account for the energy loss; a second order
polynomial in $M_{\ee}$ was used to describe the background.
The processes contributing
to the background are described in section 6.3.
In this fit, the fraction of events undergoing
bremsstrahlung was constrained to be that
determined from the Monte Carlo study described below.
The mass (\mbox{$\rm 3.08 \pm 0.01 ~GeV$})
and the resolution (\mbox{$\rm 38 \pm 10~ MeV$}), used as free parameters
in the fit, were in agreement
with the predictions of the Monte Carlo simulation. The fit,
shown in Fig.~\ref{effwgp}a, yields $72 \pm 9$ events for the \jpsi~signal.

$\bullet$ Muon decay mode:
the muon data sample still contained some contamination from
cosmic ray muons. These were identified and removed by applying a
collinearity criterion. A total of 45 events
satisfied the selection criteria in the invariant mass range
between 2 and
4 GeV, with 35 events in the invariant mass
range between 2.85 and $\rm 3.25~ GeV$. Figure~\ref{effwgp}b
shows the invariant mass
spectrum: a clear peak at the \jpsi~mass is observed.
The mass spectrum was fitted with the combination of a Gaussian and a flat
background, shown in Fig.~\ref{effwgp}b.
%The background is shown by the dotted line.
The fitted mass (\mbox{$\rm 3.08 \pm 0.01 ~GeV$}) and the resolution
(\mbox{$\rm 61 \pm 13~ MeV$}) were again in agreement with the Monte Carlo
expectations. The fit yielded
$32 \pm 6$ signal events.

\subsection{\bf Monte Carlo simulation and acceptance}

Elastic \jpsi~production was simulated with the
DIPSI\cite{DIPSI} and the EPJPSI\cite{EPJPsi}~generators.
The model used by DIPSI
assumes that the exchanged photon fluctuates into a $\rm q\bar q$
pair which then interacts with the proton via the exchange of a
pomeron described in terms of a gluon ladder\cite{RYSKIN}.
The model is based on a perturbative QCD calculation in the leading log
approximation. The hard scale
in this model is taken to be $\mu^2 \simeq M_{\rm{J}/\psi}^2/4
\simeq 2.5$ GeV$^2$.
The gluon momentum density
of the proton,
used as input in the DIPSI Monte Carlo, was $\propto x^{-0.4}$.
The EPJPSI generator assumes pomeron exchange for the
elastic \jpsi~production.
Two different models of the pomeron were used.
In the first model, the
pomeron consists of two gluons, and one of the gluons interacts with
the \ccbar~state into which the photon has fluctuated;
the \jpsi~is formed from
the \ccbar~and the remaining gluon of the pomeron.
In the second model, the pomeron is without a structure, and
it interacts with the photon as a whole to form the \jpsi.

Events were generated in the $W$ range between 30 and 200 GeV and between
the minimum allowed value of \qsquare~($\simeq 10^{-10}$ GeV$^2$)
and 4 GeV$^2$.
The events were then passed through the standard ZEUS
detector and trigger simulation programs, and processed with the same
reconstruction and analysis programs as the data.
The distributions of the reconstructed kinematic quantities obtained
using DIPSI were in good agreement with those from the data.
The overall acceptance (including the
geometric acceptance, detector, trigger and reconstruction efficiencies)
was then obtained using DIPSI as the ratio of the number of accepted
Monte Carlo events to the number generated in
the selected kinematic range of W between 40 and 140 GeV. Events generated
with EPJPSI reproduced some aspects of the data well and
were used in the study of systematic uncertainties in the acceptance
determination. Table 1 shows
the acceptances\footnote{The efficiency of the
CAL trigger threshold for the muon decay mode was found to be 90\%; this
has been included in the acceptance values. Also included is an additional
factor of 0.9 for the muon chamber efficiency.}
in various $W$ ranges determined for each decay mode.

\subsection{\bf Background}

Two types of background contributions were considered: the first is a
continuum background
present over the complete mass range, which was already
subtracted in obtaining the number of signal events from the fit;
the second type is \jpsi~production through
proton dissociation or other inelastic production processes.

$\bullet$ Continuum background: the
Bethe-Heitler process produces lepton pairs from photon-photon scattering
where the electron and the proton each radiates a photon.
The invariant mass
spectrum of the lepton pair, either \ee or \mm, typically forms a
continuum with a maximum at low masses.
The contribution from this process was obtained by
generating Monte Carlo events using the generator LPAIR\cite{LPAIR}.
These events were treated the same way as those generated by DIPSI.
The hatched regions in Fig.~\ref{effwgp}a and~\ref{effwgp}b
show the LPAIR events,
normalized
to the appropriate ep luminosity, which survived all the selection criteria.
In the mass region between 2.85 and 3.25 GeV, the background from this process
in each mode is $\sim 8.5\%$.

For the electron decay mode, a
second source of continuum background was responsible for the difference
between the quadratic polynomial (dashed curve)
and the two photon process shown in Fig.~\ref{effwgp}a
(hatched area).
After studies with data and Monte Carlo,
this was determined to be from pion contamination.
One or both of the electron candidates could be misidentified pions;
the probability
of misidentification decreases with increasing momentum. In the mass range
between 2 and 4 GeV, the background from the pion contamination
is comparable to that from the two-photon process.

$\bullet$ the proton dissociation process in
reactions like ep $\rightarrow$ e~\jpsi~$X$, where $X$ was undetected,
can also contribute to the observed \jpsi~signal,
as the outgoing proton is not observed.
This background was estimated by removing the criterion that the
energy deposited in CAL, excluding that from the \jpsi~decay leptons,
be less than 1 GeV.
A total of four events in the electron mode and three events in the muon mode
were observed with CAL close to the proton direction; these
are candidates for \jpsi~production accompanied by proton dissociation.
Monte Carlo events simulating this process were generated with
PYTHIA\cite{PYTHIA}; the mass distribution of the diffractive system $X$
was parametrised as
$d\sigma/d{M_X}^2 \sim {M_X}^{-n}$ where $n$ was varied\footnote
{The CDF Collaboration reported a measurement of
$n=2.20 \pm 0.03$ at $\sqrt s = 1800$ GeV\cite{DIFMON},
in agreement with Regge
theory predictions\cite{THMON}, which is well within the range between 2 and 3
considered.}
between 2 and 3.
These Monte Carlo events were analysed in the same way as the data.
For $n=2.5$, the proton dissociative contamination in the \jpsi~signal
was estimated to be $17\%$, by comparing
the number of events with extra energy in CAL from this Monte Carlo sample
with the seven events from the data.
The fraction of proton dissociative contamination determined was
(17$^{+8}_{-5} \pm 10)\%$, where the first uncertainty is
statistical, obtained from the electron and the muon modes together,
and the second uncertainty is systematic, observed from the variation
of $n$ between 2 and 3 in the Monte Carlo.
The cross section for the elastic process was thus obtained by
subtracting 17\% from the \jpsi~signal, independent of $W$.

$\bullet$ the contribution from the photon-gluon fusion process
was determined from Monte Carlo studies using HERWIG\cite{HERWig}~and
EPJPSI\cite{EPJPsi}. Possible
contributions from $\psi(3685)$ production and processes
where the photon undergoes diffractive dissociation e.g.,
$\gamma \rightarrow \jpsi~X$, where
$X$ was not detected, were also considered.
The total contribution from these processes was found to be negligible and
a systematic uncertainty of 3\% was assigned to it.

\section{\bf Results}
\subsection{\bf ep cross section}

The electroproduction cross section, $\rm \sigma_{ep}$, is calculated as:
$ \rm \sigma_{ep}$ = ${1 \over A} \times
{ 1 \over Br} \times { 1 \over {\cal L}}
\times N,  $
where $N$ denotes the number of \jpsi~signal events, $A$ the acceptance,
$\cal L$ the integrated luminosity, and $Br$ the leptonic branching
fraction of \jpsi\cite{Pdata},
namely (5.99 \ppm~0.25)\% for \ee and (5.97 \ppm~0.25)\% for \mm.
%The results of the cross section measurements in
%the complete W range of 40 GeV$< W <140$ GeV, without subtracting the
%proton diffractive dissociative contribution, are :
%$\rm \sigma_{ep} = 8.0 \pm 1.0 ~nb$, and
%$\rm \sigma_{ep} = 7.3 \pm 1.4 ~nb$, from the electron and the muon decay
%modes, respectively. The uncertainties are statistical only.
Subtracting the 17\% contribution to the \jpsi~signal from the proton
dissociation process, we obtain for the elastic cross section for
the process
ep $\rightarrow$ e~\jpsi ~p for $\qsquare < 4$ GeV$^2$ and
$\sqrt s = 296$ GeV
in the $W$ range between 40 and 140 GeV:
$\rm \sigma_{ep} = 6.6^{+1.0}_{-1.3} ~nb$, and
$\rm \sigma_{ep} = 6.2^{+1.1}_{-1.6} ~nb$, from the electron and the muon
decay modes, respectively. The errors are statistical.

\begin{figure}[hptb]
\unitlength1cm  \begin{picture}(8,15.0)
\includegraphics{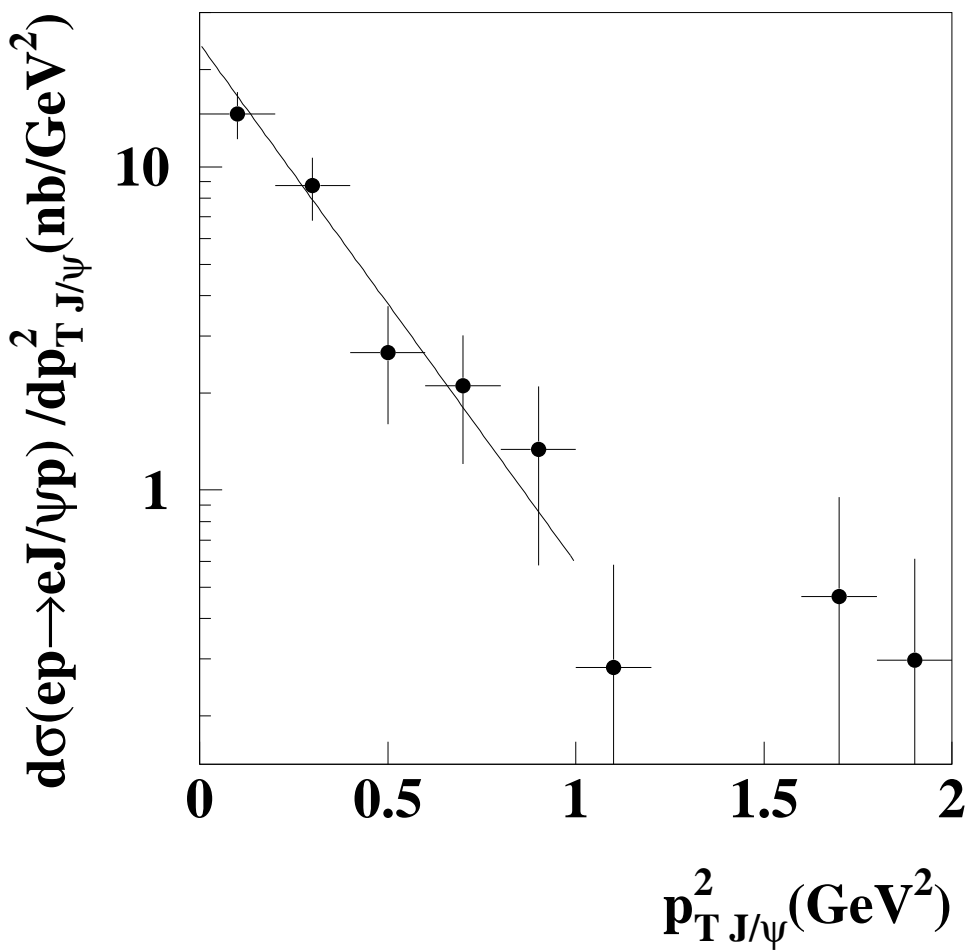}
\end{picture}
\caption{
the differential cross section $d\sigma_{ep}/dp_{T\jpsi}^2$ for
e$^-$p $\rightarrow$ e$^- \rm \jpsi$ p
from both decay channels combined for
40 $\le W \le$ 140 GeV. The continuous line indicates the result of a fit
with the function exp($-b\ptsq_{\jpsi}$) with
$b$ = 3.7$\pm 1.0$ GeV$^{-2}$. Only statistical errors are shown.}
\label{effplt}
\vspace{-0.2cm}
\end{figure}

Figure~\ref{effplt}~shows the d$\rm \sigma_{ep}$/d$p_{T\rm{J}/\psi}^2$
differential
cross section for both decay channels combined
after the background subtractions.
An exponential fit of the form exp($-b\ptsq_{\rm{J}/\psi}$)
to the distribution in the range
$0\leq \ptsq_{\rm{J}/\psi} \leq 1$ GeV$^2$ yields a slope $b$ of
$ 3.7 \pm 1.0$ GeV$^{-2}$,
while a fit in the range $0\leq \ptsq_{\rm{J}/\psi} \leq 0.75$ GeV$^2$
gives a slope $b$ of $ 4.5 \pm 1.4$ GeV$^{-2}$,
where the statistical and the systematic
errors have been added in quadrature. The acceptance
in these $\ptsq_{\rm{J}/\psi}$ ranges was constant within $15\%$.
As noted earlier, for $\qsquare\simeq 0$, $\ptsq_{\rm{J}/\psi}$
approximates $\vert t \vert $. Taking into account the \qsquare~dependence,
the slope of the $\vert t \vert$ distribution obtained from the Monte Carlo
is $\sim 0.5$ unit higher than the slope obtained from
the $\ptsq_{\rm{J}/\psi}$ distribution, in the $\ptsq_{\rm{J}/\psi}$ ranges
described above.

In order to determine the cross section dependence on $W$, the data samples
were divided into two $W$ ranges: $40 \leq \ W\  \leq 90$ GeV and
$90 \leq \ W\ \leq 140$ GeV.
Table 1 lists the elastic cross sections from each $W$ range
after subtracting the backgrounds, along with
the numbers of events and the acceptances.

\begin{center}
{\bf Table 1: Acceptance and cross sections} \\
\vspace{0.3 cm}
\begin{footnotesize}
\begin{tabular}{|c||c|c|c|c|}
\hline
W range (GeV) & \multicolumn{2}{|c|}{ 40 - 90 } &
\multicolumn{2}{|c|}{ 90 - 140 }  \\
\hline\hline
decay mode &\jpsitoee & \jpsitomm & \jpsitoee & \jpsitomm \\ \hline
acceptance & 0.39 & 0.11 & 0.22 & 0.19  \\ \hline
signal events & 44$^{+7}_{-11}$ & 15$^{+4}_{-5}$ &
      16 \ppm~4 & 12$^{+5}_{-6}$ \\ \hline
$\rm \sigma_{ep}$(nb)& 3.9$^{+0.6}_{-1.0}$\ppm0.7&4.7$^{+1.2+0.9}_{-1.6-1.0}$&
 2.5\ppm0.6\ppm0.4  &  2.2$^{+0.9+0.4}_{-1.0-0.5}$ \\ \hline \hline
integrated photon flux & \multicolumn{2}{|c|}{ 0.077
}&\multicolumn{2}{|c|}{0.033} \\ \hline
$\rm \sigma_{\gamma p}$(nb) & 50$^{+8}_{-13}$\ppm~10 & 61$^{+16+14}_{-20-14}$&
76$^{+19}_{-24}$ \ppm~13 & 65$^{+16+13}_{-33-14}$ \\ \hline \hline
$\langle W \rangle$~(GeV)  & \multicolumn{2}{|c|}{67\ppm~11
}&\multicolumn{2}{|c|}{114 \ppm~9} \\ \hline \hline
$\sigma_{\gamma p}$~(nb)   & \multicolumn{2}{|c|}{52$^{+7}_{-12}\pm 10$} &
\multicolumn{2}{|c|}{71$^{+13}_{-20}\pm 12$}  \\ \hline
\end{tabular}
\end{footnotesize}
\end{center}

%$\langle W \rangle$(GeV)& 61 \ppm 9 & 58 \ppm 14 & 106 \ppm 28 & 114 \ppm 11
%%\\ \hline

\subsection{Systematic uncertainties}
The summary of the uncertainties from various sources is reported in Table 2.
The polarization of the \jpsi~was not measured and
the angular distribution of the decay leptons was varied
from flat to the form $1+\cos^2 \theta^*$, where $\theta^*$ is the decay angle
of
the leptons in the \jpsi~rest frame with respect to the \jpsi~lab momentum.
This could affect the acceptance by
up to $5 \%$. No uncertainty was attributed to the $W$ calculation, as the
resolution in $W$ is of the order of 1\%.
The variation in acceptance from the modeling of the $W$ dependence
was calculated using the different Monte Carlo
generators described in section 6.2; an uncertainty of 9\% was
assigned to each of the decay modes.
The uncertainty in acceptance from the variation in the gluon density
was included in modeling the $W$ distribution in the Monte Carlo simulations.
The computation of the muon chamber efficiency added an asymmetric
uncertainty as shown in Table 2.
The uncertainty from the track multiplicity determination was
observed to be 5\%. The energy deposit in CAL was obtained after the
uranium noise subtraction\cite{CAL}. This led to a 4\% uncertainty.
The uncertainty due to the calorimeter trigger thresholds for the
muon decay channel was 10\%.
For the electron mode, the energy requirement
for triggering was far enough above the threshold that only a maximum of
5\% variation could be observed in the efficiency determination.

An uncertainty of 10\% was attributed to the subtraction of the
proton dissociation process, and a 3\% uncertainty was added
to account for contributions from other possible inelastic processes
to the signal, referred to as feed-in from
other modes, as discussed in section 6.3.
A total uncertainty of
17\% for the electron mode and $^{+20\%}_{-21\%}$ for the muon mode in the
electroproduction cross section was
thus obtained by adding all contributions in quadrature.
\newpage
\begin{center}
{\bf Table 2: Systematic uncertainties for the
{\boldmath ep $\rightarrow$ e~\jpsi~p} \\
cross section measurement} \\
\vspace{0.3 cm}
\begin{footnotesize}
\begin{tabular}{|l|r|r|}
\hline
uncertainties  & \jpsitoee 	& \jpsitomm	\\
\hline \hline
branching fraction                            &4 \% & 4 \% \\
luminosity 	                                  & 3.3 \% & 3.3 \% \\
\jpsi~decay angular distribution   	             & 5 \% & 5 \% \\
W dependence                                         & 9 \%& 9 \% \\
muon chamber efficiency                     &   & +5 \% $-$9 \% \\
track multiplicity                           & 5 \% & 5 \%\\
$\leq$1 GeV energy requirement          & 4 \% & 4 \% \\
CAL trigger threshold                   & 5 \%  & 10 \% \\
proton dissociation subtraction                   & 10 \% & 10 \% \\
feed-in from other modes 	                & 3 \% & 3 \% \\ \hline
   total 	                            & 17 \% & +20 \% $-$21 \% \\ \hline
\end{tabular}
\end{footnotesize}
\end{center}

\subsection{\bf Photoproduction cross section}

The photon-proton cross section
$\sigma_{\gamma p \rightarrow {\rm J}/\psi p}$ is obtained from
the corresponding electron-proton
cross section by using the relation :
%\subsection{\mbox{\boldmath $\gamma p$} Cross Section }
\[
\sigma_{ep}(s) = \int_{y_{min}}^{y_{max}} dy \int_{Q^2_{min}}^{Q^2_{max}} dQ^2
\,\cdot \Phi(y,Q^2) \,
 \cdot \sigma_{\gamma^* p}(y,Q^2) ,
\]
where
\[
\Phi(y,Q^2) = \frac{\alpha}{2 \pi} \frac{1}{y Q^2}  [1+(1-y)^2- \frac{2 m_e^2
y^2}{Q^2}]
\]
is the photon flux factor\cite{EPA},
$Q^2_{min}=m_e^2 \frac{y^2}{1-y}$,
$Q^2_{max} =4$ GeV$^2$ and $m_e$ is the electron mass.
Since the median \mbox{$Q^2 \approx 10^{-3}~\rm GeV^2$} is very small,
we can neglect the longitudinal contribution and the $Q^2$
dependence of $\sigma_{\gamma^* p}$.
The $\gamma p$ cross section is then obtained as the ratio of
the measured
$e p$ cross section and
the photon flux factor $\Phi$ integrated over the $Q^2$ and $y$ range
covered by the measurement.
This procedure assumes that $\sigma_{\gamma p}$ is independent of $y$
in the range of the measurement.
As this dependence is not known a priori,
the above calculation was repeated assuming a rise of
$\sigma_{\gamma p}(W)$ proportional to $W~(=\sqrt {sy})$.
An increase of 10\% in the resulting cross section
was found at $\langle W \rangle = 67$ GeV
and less than 2\% at $\langle W \rangle = 114$ GeV. These have
been added in the systematic uncertainty in the photoproduction cross section
measurements. The cross sections and the integrated photon flux in the
different $W$ ranges are summarized in Table 1.
The first uncertainty quoted for the cross sections is statistical
and the second is systematic.
Combining results from the two leptonic decay modes, the measured
\jpsi~photoproduction elastic cross sections are :
\begin{eqnarray*}
\sigma_{\gamma p \rightarrow \rm{J}/\psi p} & = & 52^{ \ +7}_{-12}\pm 10~nb \ \
for~\langle W \rangle = 67~\rm GeV~and \\
\sigma_{\gamma p \rightarrow \rm{J}/\psi p} & = & 71^{+13}_{-20}\pm 12~nb \ \
for~\langle W \rangle = 114~ \rm GeV.
\end{eqnarray*}
%\vglue0.15cm
The first error is the weighted error of the uncertainty specific to each decay
mode where this uncertainty is obtained by combining the statistical error and
the systematic error unique to the decay mode in quadrature; this
error was used to
obtain the weighted combined cross section. The second error is the systematic
error common to both decay
modes.

\section{\bf Discussion}

\begin{figure}[hptb]
\unitlength1cm  \begin{picture}(8,15.0)
\includegraphics{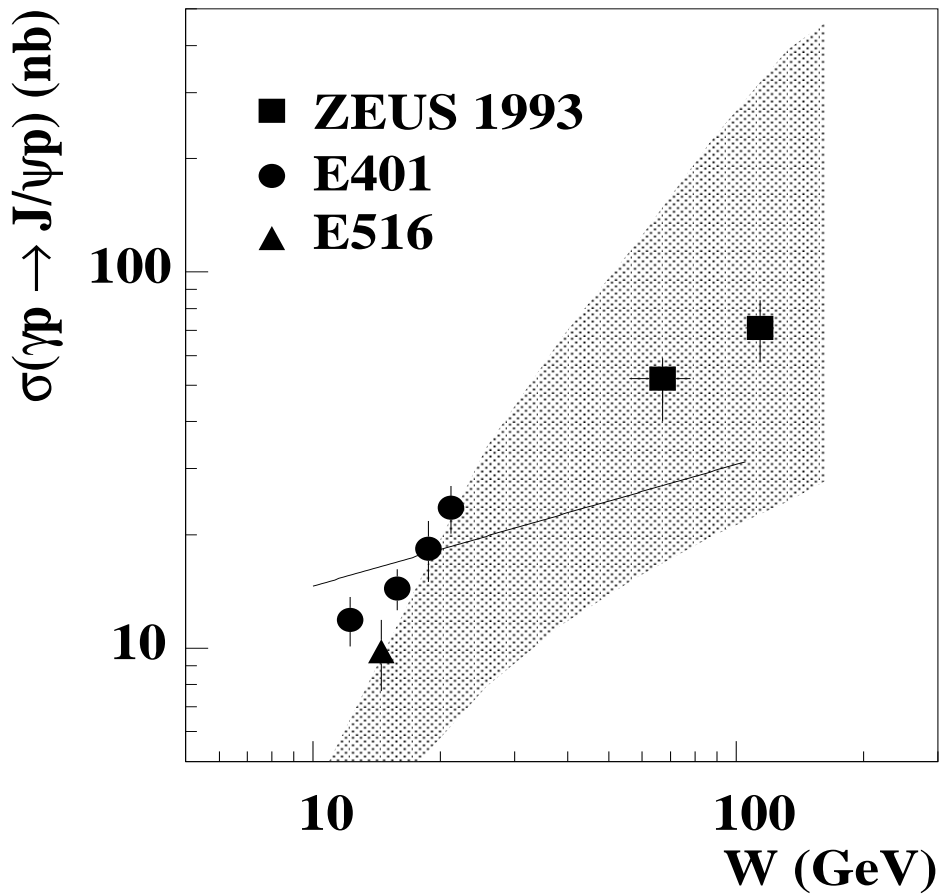}
\end{picture}
\caption{A compilation of \jpsi~elastic cross sections in photoproduction. The
solid squares represent the measurements from the ZEUS experiment
at $\langle W \rangle$ values of 67
GeV and 114 GeV. The shaded region represents the prediction of the
Ryskin model{[5]} using the upper
and lower limits of the gluon momentum
density as extracted by the ZEUS experiment at $Q^2=7$ GeV$^2${[25]} in LO
and evolved back to 2.5 GeV$^2$. The
solid line is a VDM-like prediction{[24]}.}
\label{xsect}
\vspace{-0.2cm}
\end{figure}

Figure~\ref{xsect} displays a compilation
of the \jpsi~elastic photoproduction
cross section measurements\cite{AL_xsect2}. The compilation
includes only results from the fixed target experiments which measured
the recoil proton.
The elastic \jpsi~photoproduction measurements by ZEUS are also
shown, where the elastic cross section has been obtained by subtracting
the proton dissociative production of the \jpsi.
A significant rise of the \jpsi~cross section with the c.m. energy is visible.
%The measurement of the H1
%collaboration\cite{H1M}~of the \jpsi~cross section
%at $\langle W \rangle$ = 90 GeV
%is $56 \pm 13 \pm 14$ nb for a total of 26 signal events observed
%in the \ee~and the \mm~modes combined.
%However, it
%includes an unknown fraction of events
%from the reaction $\gamma p \rightarrow \jpsi~+ X$,
%with no extra track between $\rm \theta = 7^{\circ} ~and ~170^{\circ}$.
%Hence, this
%measurement is not presented in Fig.\ref{xsect}.

Theoretical predictions based on Regge-type
and QCD inspired models are also shown in Fig.~\ref{xsect}.
The solid line is the prediction of
Donnachie and Landshoff, normalized to lower energy data, using a
supercritical pomeron\cite{NEWDL}.
%The dotted curve is a calculation by Gotsman, Levin
%and Maor\cite{GLM12}, who add screening corrections to the previous
%model\cite{DANDL}.
It predicts a slower rise in
cross section with $W$ than is observed.
This is also true of other VDM type models\cite{GLM12,TERRON}.
The essential point of the QCD inspired models is that the cross section
is proportional to the square of the gluon density.
At the HERA energy ranges very low
$x$ values (from $\sim 5\times 10^{-4} \rm \ to \ 5\times 10^{-3}$ in the
present analysis) contribute to \jpsi~production.
The shaded band in Fig.~\ref{xsect} is the
prediction of the Ryskin model\cite{RYSKIN},
obtained with DIPSI (with $\alpha_s = 0.25$),  using the
upper and lower limits of the leading order (LO) gluon momentum density
as extracted by the ZEUS experiment\cite{ZGLUE}~from the scaling
violation of $F_2$ at \qsquare~= 7 GeV$^2$
and evolved back to \qsquare~= 2.5 GeV$^2$,
the scale used in the model.
The energy
behaviour shown by the shaded band is in accord with the data.
A recent modification by Ryskin\cite{NWRYS}, which uses
%some saturation effects\cite{NWRYS}, and that
a calculation based on the present model\cite{RYSKIN} and a gluon distribution
with some saturation effects is also in agreement with our
measurements; so is the calculation
of Nemchik, Nikolaev and Zakharov\cite{NIKZAK}, which uses the
dipole cross section
solution of the generalized BFKL \cite{BFKL} equation.
%These QCD models are also
%which predicts an
%elastic \jpsi~cross section of 90 nb at W = 200 GeV,
%in good agreement with our measurements.

\section{\bf Conclusions}

We have measured the elastic photoproduction cross section of \jpsi~in ep
interactions at $\gamma p$ c.m. energies between 40 and 90 GeV
($\langle W \rangle$ = 67 GeV) and
90 and 140 GeV ($\langle W \rangle$ = 114 GeV).
The \jpsi~was detected in its leptonic decay modes
(\ee~or \mm) in events where the scattered electron and
proton were not observed.
The elastic photoproduction cross sections, obtained from the
combined electron and muon decay modes, are :
\begin{eqnarray*}
\sigma_{\gamma p \rightarrow \rm{J}/\psi p} & = & 52^{ \ +7}_{-12}\pm 10~nb \ \
for~\langle W \rangle = 67~\rm GeV~and \\
\sigma_{\gamma p \rightarrow \rm{J}/\psi p} & = & 71^{+13}_{-20}\pm 12~nb \ \
for~\langle W \rangle = 114~ \rm GeV.
\end{eqnarray*}
The observed rise in the cross section compared to the lower energy
measurements is not adequately described
by Regge-type models and is better represented by perturbative calculations
if a rise in the gluon density in the proton at low $x$ is assumed.

\section{Acknowledgement}

We thank the DESY Directorate for their
strong support and encouragement. The remarkable achievements of the HERA
machine group were essential for the successful completion of this work,
and are very much appreciated. We also gratefully acknowledge the support
of the DESY computing and network group. Finally, we wish to thank
N.~N.~Nikolaev and M.~Ryskin for useful discussions.
%%
%       Reference
%

%

\newpage
%\section{\bf Figures}

\newpage

\newpage

\newpage

\end{document}